\documentstyle[epsfig]{mn}

\newif\ifAMStwofonts

\def\etal{{\it et al.} }
\def\astpart{{\it Astrop. Physics}}

\def\apj{{\it ApJ}}
\def\apjl{{\it ApJ Lett.}}

\def\aas{{\it Astron.\ Astrophys.\ Suppl.}}
\def\aa{{\it A\&A}}

\def\newa{{\it New Astr.}}
\def\nature{{\it Nature}}

\def\inpress{{\it in press}}

\def\mic{{{\mu}m}}
\def\uk{{{\mu}K}}
\def\cm2{$cm^{-2}$}

\ifoldfss
  \ifCUPmtlplainloaded \else
    \NewTextAlphabet{textbfit} {cmbxti10} {}
    \NewTextAlphabet{textbfss} {cmssbx10} {}
    \NewMathAlphabet{mathbfit} {cmbxti10} {} 
    \NewMathAlphabet{mathbfss} {cmssbx10} {} 
  \fi
  \ifAMStwofonts
    \ifCUPmtlplainloaded \else
      \NewSymbolFont{upmath} {eurm10}
      \NewSymbolFont{AMSa} {msam10}
      \NewMathSymbol{\upi}     {0}{upmath}{19}
      \NewMathSymbol{\umu}     {0}{upmath}{16}
      \NewMathSymbol{\upartial}{0}{upmath}{40}
      \NewMathSymbol{\leqslant}{3}{AMSa}{36}
      \NewMathSymbol{\geqslant}{3}{AMSa}{3E}

    \fi
  \fi
\fi 
\ifnfssone
  \newmathalphabet{\mathit}
  \addtoversion{normal}{\mathit}{cmr}{m}{it}
  \addtoversion{bold}{\mathit}{cmr}{bx}{it}
  \newmathalphabet{\mathbfit} 
  \addtoversion{normal}{\mathbfit}{cmr}{bx}{it}
  \addtoversion{bold}{\mathbfit}{cmr}{bx}{it}
  \newmathalphabet{\mathbfss} 
  \addtoversion{normal}{\mathbfss}{cmss}{bx}{n}
  \addtoversion{bold}{\mathbfss}{cmss}{bx}{n}
  \ifAMStwofonts
    \ifCUPmtlplainloaded \else
      \UseAMStwoboldmath
      \makeatletter
      \new@mathgroup\upmath@group
      \define@mathgroup\mv@normal\upmath@group{eur}{m}{n}
      \define@mathgroup\mv@bold\upmath@group{eur}{b}{n}
      \edef\UPM{\hexnumber\upmath@group}
      \new@mathgroup\amsa@group
      \define@mathgroup\mv@normal\amsa@group{msa}{m}{n}
      \define@mathgroup\mv@bold\amsa@group{msa}{m}{n}
      \edef\AMSa{\hexnumber\amsa@group}
      \makeatother
      \mathchardef\upi="0\UPM19
      \mathchardef\umu="0\UPM16
      \mathchardef\upartial="0\UPM40
      \mathchardef\leqslant="3\AMSa36
      \mathchardef\geqslant="3\AMSa3E
    \fi
  \fi
\fi 

\ifnfsstwo
  \DeclareMathAlphabet{\mathbfit}{OT1}{cmr}{bx}{it}
  \SetMathAlphabet\mathbfit{bold}{OT1}{cmr}{bx}{it}
  \DeclareMathAlphabet{\mathbfss}{OT1}{cmss}{bx}{n}
  \SetMathAlphabet\mathbfss{bold}{OT1}{cmss}{bx}{n}
  \ifAMStwofonts
    \ifCUPmtlplainloaded \else
      \DeclareSymbolFont{UPM}{U}{eur}{m}{n}
      \SetSymbolFont{UPM}{bold}{U}{eur}{b}{n}
      \DeclareSymbolFont{AMSa}{U}{msa}{m}{n}
      \DeclareMathSymbol{\upi}{0}{UPM}{"19}
      \DeclareMathSymbol{\umu}{0}{UPM}{"16}
      \DeclareMathSymbol{\upartial}{0}{UPM}{"40}
      \DeclareMathSymbol{\leqslant}{3}{AMSa}{"36}
      \DeclareMathSymbol{\geqslant}{3}{AMSa}{"3E}
    \fi
  \fi
\fi 

\ifCUPmtlplainloaded \else
  \ifAMStwofonts \else 
    \def\upi{\pi}
    \def\umu{\mu}
    \def\upartial{\partial}
  \fi
\fi

\title{Map-making methods for Cosmic Microwave Background experiments}
\author[X. Dupac \& M. Giard]
{Xavier Dupac \& Martin Giard\\
Centre d'\'Etude Spatiale des Rayonnements (CESR)\\ 
9 av. du colonel Roche, BP4346, F-31028 Toulouse cedex 4, France}

\date{Accepted
      Received
      in original form}

\pagerange{ -- }
\pubyear{2001}

\begin{document}

\maketitle

\begin{abstract}
The map-making step of Cosmic Microwave Background data analysis involves
linear inversion problems which cannot be performed by a brute force approach
for the large timelines of today.
We present in this article optimal vector-only map-making methods, which are
an iterative COBE method, a Wiener direct filter and a Wiener iterative
method.
We apply these methods on diverse simulated data, and we show that they
produce very well restored maps, by removing nearly completely the correlated
noise which appears as intense stripes on the simply pixel-averaged maps.
The COBE iterative method can be applied to any signals, assuming the
stationarity of the noise in the timeline.
The Wiener methods assume both the stationarity of the noise and the sky,
which is the case for CMB-only data.
We apply the methods to Galactic signals too, and test them on balloon-borne
experiment strategies and on a satellite whole sky survey.
\end{abstract}

\begin{keywords}
cosmology: Cosmic Microwave Background -- methods: data analysis.
\end{keywords}

\section{Introduction}
The Cosmic Microwave Background (CMB hereafter) is being extensively studied nowadays,
thanks to the improvement of instrument and detector
performances.
Since the COBE experiment \cite{smoot92}, which performed the first detection of CMB
anisotropies, the size of time-ordered information (TOI hereafter) has been widely increased.
The data processing and analysis is thus still a
challenge for large timeline data, already in processing or still to
come.
MAP (see for example Wright 1999) and Planck (see for example Tauber 2000) experiments will produce huge timelines of
hundreds of million samples or more, and even the balloon-borne experiments, BOOMERanG \cite{debernardis00}, MAXIMA \cite{hanany00},
Archeops \cite{benoit01} and
many others, produce large timelines for many channels.
The Cosmic Microwave Background data analysis is usually performed
in three steps, the first being the map-making process, the second the $C_{l}$
estimation from the maps and the noise covariance matrices, the third the
cosmological parameters estimation from the $C_{l}$ power spectrum.
In this article, we focus on the map-making step, in the context of
computing difficulties due to large timelines.
We also explore how the scanning strategy
influences the map-making process and the final amount of noise in the maps.
Map-making, in the context of large timelines, cannot be performed by a
brute-force approach, which would imply the manipulation and inversion of
tera-element large matrices.
Non optimal map-making methods have been developed, such as destriping using the scan
intercepts (see Delabrouille 1998 and Dupac \& Giard 2001).
We aim in this paper to apply the optimal methods to large timelines,
developing algorithms to avoid computation trouble.
The map-making methods for CMB are known to be optimal when following
desirable properties, such as minimizing the reconstruction error and the
noise chi square towards the reconstructed map, and being the maximum likelihood solution for
simple conditions or no conditions.
These are linear map-making methods, well known as the COBE method \cite{janssen92} and
the Wiener filter \cite{wiener49}.
The linear methods allow to find an analytical expression for such
requirements: the time-ordered data, represented
by a vector y, are a linear expression of the real sky, represented
by a vector x.
(We note every vector with a small letter and every matrix with a capital letter.)
We have thus:

$y = A x + n$ \hspace*{5em} (1)

where A is the point-spread (convolution) matrix and n the noise vector
in the timeline.
For CMB experiments, x would represent the pixelised sky map of the CMB temperature.
The map-making problem is thus written as this:

\~x $= W y$ \hspace*{6em} (2)

where \~x is the vector of the reconstructed sky map, and W the inversion
matrix.
The COBE method can be derived by minimizing the noise chi square towards the
reconstructed map, the reconstruction error subject to the constraint W A = I,
and is the maximum likelihood solution for the sky if the noise is gaussian
(no prior on the sky) (see for example Tegmark 1997).
It is written as:

$W = [A^{t} N^{-1} A]^{-1} A^{t} N^{-1}$ \hspace*{5em} (Eq. 3, COBE)

where N is the covariance matrix of the noise in the timeline: N = $<nn^t>$.
The Wiener filter can be derived as the solution of the absolute minimization
of the reconstruction error \cite{bunn94}, or as the maximum probability
solution for the sky if both the noise and the sky are {\it a priori} gaussian
\cite{zaroubi95}.
It is written as:

$W = [S^{-1} + A^{t} N^{-1} A]^{-1} A^{t} N^{-1}$ \hspace*{2em} (Eq. 4, Wiener 2)

or:

$W = S A^{t} [A S A^{t} + N]^{-1}$ \hspace*{5em} (Eq. 5, Wiener 1)

where S is the covariance matrix of the sky in the map domain: S = $<xx^t>$.
These two writings are the same matrix but can be totally
different toward algorithmics. 
From these theoretically optimal methods, one can imagine other
slightly different methods, made to solve such or such practical
problem (see Tegmark 1997 for a large set of linear and non linear
methods).
The COBE method (Eq. 3) and the Wiener filter (Eq. 4, 5) are both known to retain all the
cosmological information present in the TOI \cite{tegmark97}.
These methods are optimal subject to the condition
that the noise is uncorrelated with the sky, which is a good assumption for
low contrasts maps such as CMB ones.
One finds this assumption in every paper concerning CMB map-making, and it is
not the purpose of this article to question this, however, following the objects observed and the nature of the
observation strategy, the bolometer response could be less perfect, and
therefore induce some noise correlated to the sky signal.

Applying these optimal methods to large timelines is not
straightforward, because of computer limitations.
The several tens or hundreds of mega-elements in a bolometer timeline
have to be processed by vector-only methods, that we aim to present in this
paper.
We mean by "vector-only methods" computations in which one never needs to
compute or even to store any large matrix such as A, N or S.
In this context, the way to make the timeline, that is the scanning
strategy, can be important towards map-making and the quality of the final products.
We present in Section 2 diverse balloon and satellite scanning
strategies for CMB experiments. These aim to be representative of the
different priorities one aims to focus on: redundancy, coverage, whole
sky survey..., and how these strategies influence the map-making efficiency
and the final quality of the retrieved maps.
In Section 3 we present how we simulate our timelines.
In Section 4 we present the map-making methods, applied to large
timelines, and discuss their conditions of application.
In Section 5 we compare the results on the reconstructed maps, for the different map-making methods and experiment
strategies that we have investigated.

\section{Observation strategy}
We present here three different large-coverage observation strategies,
simulated from simple and usual technical requirements.
We simulate balloon-borne CMB experiments, required to scan the sky making
constant elevation circles at a rather constant rotation speed (as do
Archeops and TopHat: http://topweb.gsfc.nasa.gov).
This allows to observe a large area on the sky, thanks to the rotation of the
Earth (and, eventually, to the moving of the balloon on the Earth). In this
context, it is easy to understand that the launching place of the balloon and
the flight duration will have a crucial influence on the sky coverage and the
redundancy in the map pixels.
Experiments using this constant elevation scanning strategy have to
balance between coverage and redundancy. Maximum redundancy is obtained
with high elevation angles and high latitude launch places. To the
contrary, the best coverage over flight time ratio is obtained at
 near-equatorial places and with low elevation angles.
The factors controlling the observation characteristics are the place and date
of launch, the trajectory of
the balloon, the flight duration, the elevation angle of scanning above the
horizon, the sampling frequency, and the rotation speed of the gondola
(especially if it is constant or not).
We have simulated two balloon-borne experiment timelines with a constant
scanning elevation of 35 degrees above the horizon, a sampling frequency of
100 Hz and a rotation speed of the gondola of 2 rpm.

\begin{table}
\caption[]{\label{table: experiments}
Caracteristics of the three observing strategies.
}
\begin{flushleft}
\begin{tabular}{llll}
\hline
 & Polar & Equatorial & Satellite\\
\hline
\hline
Observing time & 24 h & 12 h & {\small (under-sampling)}\\
\hline
Coverage & 35 \%  & 61 \% & 100 \%\\
\hline
Number of samples & 8 & 4 & 8\\
(millions) & & & \\
\hline
Avg redundancy & 30.5 & 8.7 & 11.8\\
per pixel & & & \\
\hline
\end{tabular}
\end{flushleft}
\end{table}

One simulated experiment is designed to maximize the redundancy per pixel,
however covering a large part of the sky.
It is a 24 hours polar flight from Kiruna, Swedish Lappland.
The winter time in this region provides polar nights allowing to make 24 hours
flight avoiding contamination from the sun.
The coverage for this polar flight is 35 \% of the sky.

The second simulated experiment is a 12 hours equatorial flight.
The coverage is for this 12 hours flight 61 \% of the whole sky, and would be 82 \%  for a 24
hours flight, or two 12 hours flights, one taking place six months after the first
one, to avoid sunlight.
The polar flight we present here is especially good for redundancy, while the
equatorial flight has a clearly better coverage but a much weaker redundancy.
The comparison of map-making methods on these two flights will therefore give
interesting results on how in practice the observation strategy controls the
final quality of the maps.

The third simulated scanning strategy is the one of a satellite making great circles on the sky
along ecliptic meridians.
For a satellite on a low-earth orbit (LEO), this strategy allows to always
point towards the zenith, in order to avoid contamination from the Earth.
The great circles are made perpendicularly to the sun direction, that allows
to avoid the main sunlight contamination too.
Of course, on a LEO, the exact trajectory would not be exactly as simple, because of
the gravitational perturbations that the satellite would face, but our
simulated timeline is near to what it would be.
This strategy also works for a satellite at L2, as Planck. In this case the
strategy can be as simple as a constant great circle in the ecliptic meridian
perpendicular to the Sun-Earth axis, that slowly shifts with the revolution of
the Earth. The whole sky survey is thus completed in six months.
We have simulated a quick whole sky survey made with this observing strategy,
containing 8 million samples.
Although this would not be of course the final product of a one-year satellite mission,
it allows us to compare the scanning strategy and final map
properties with those of the two balloon-borne experiments.

\section{Simulation process}
We have made our simulated skies from three components: the CMB
fluctuations, the dipole, and the Galaxy, at 2 mm wavelength.
The simulated Universe we have simulated is $\Lambda$ dominated, with
$\Omega_\Lambda = 0.7$, $\Omega_{CDM} = 0.25$, $\Omega_{bar} = 0.05$, $H_0 = 50 $
and a scalar spectral index of the fluctuations equal to 1.
The exact cosmological parameters we choose is not important for the
purpose of this article, but the gaussianity is {\it a priori} important for
the Wiener methods that we will test.
The $C_l$ simulated spectrum (with no cosmic variance) is made thanks
to the CMBFAST software \cite{seljak96}.
The sky gaussian random field (i.e. a realization of a Universe with
the cosmological parameters we chose) is then simulated with the
SYNFAST tool of the HEALPix package (http://www.eso.org/science/healpix).
The dipole is added thanks to its COBE/DMR determination
\cite{lineweaver96}, and the Galaxy at 2 mm wavelength is extrapolated from
the composite 100 $\mic$ IRAS-COBE/DIRBE all sky dataset \cite{schlegel98}.
We present in Fig. 1 a full-sky map of the Cosmic Microwave Background
fluctuations, and a full-sky 2 mm map including the
dipole and the Galaxy.
All the maps that we present in this article are Galactic Mollweide
projections centered on the Galactic center.

\begin{figure}
\begin{center}{
   \epsfxsize 10.0 true cm
    \leavevmode
  }\end{center}
\caption[]{Top: whole sky simulation of CMB fluctuations with $\Omega_\Lambda = 0.7$, $\Omega_{CDM} = 0.25$, $\Omega_{bar} = 0.05$, $H_0 = 50 $.
Bottom: whole sky simulation of CMB fluctuations and the Galaxy at 2 mm wavelength.
The maps are  Galactic Mollweide projections centered on the Galactic center.}
\end{figure}

We have simulated noisy timelines.
However, the true
time-ordered data coming directly from a true detector has not exactly
the same shape.
Usually there are glitches, systematics and noises to be subtracted before applying careful
map-making methods.
The glitches are energetic cosmic rays which hit a bolometer and produce high
and narrow peaks in the time-ordered data.
They can usually be detected and removed. 
Some noises and systematics may be removed by decorrelating with signals from
thermometers or other bolometers.
The topic of this article is not to treat these problems, however it is
important to keep in mind that the map-making that we present here is not known to
work well on brute data.
For our simulations, we aim to test our map-making methods with white and 1/f
noise.
The noise we introduce in our simulated timelines can be characterized by its
statistical power spectrum which follows a 1/f law: $l_{inf} . (1 +
(f_c/f)^n)$, where $l_{inf}$ is the level of white noise (i.e. the only noise
at high frequency), $f_c$ the cut frequency and n the power index.
These stationary noises are usually what stays in the cleaned timelines, after
removing what is possible to remove without modifying the signal.
In order to properly compare our different observing strategies, we apply the
map-making methods to timelines simulated with the same noise level and statistical properties,
with $n = 1$, $f_c = 0.1$ Hz and $l_{inf} = 100$ $\uk_{CMB}$ rms.
This white noise rms is approximatively the level expected for Planck
\cite{tauber00} bolometers with a 100 Hz sampling rate.
Of course this
quite low value is the level of the non-correlated noise in the
timelines,
but the total amount of noise introduced is much larger, as we can see in Fig. 2, which is the three simulated data
projected on a map by averaging the samples in the pixels (the simplest
map-making method, as we explain hereafter).

\begin{figure}
 \begin{center}{
   \epsfxsize 10.0 true cm
    \leavevmode
  }\end{center}\caption[]{Top: map of the polar flight data.
Center: map of the equatorial flight data.
Bottom: map of the satellite data.
The three maps contain the CMB fluctuations, the dipole, the Galaxy and the instrumental noise.)
}
\end{figure}

\section{Map-making methods applied on large timelines}
To make maps, we need first to choose a pixelization of the sphere.
We use the HEALPix scheme (http://www.eso.org/science/healpix), whose
angular resolution is defined by the parameter $N_{side}$.
We present our simulated and reconstructed maps with $N_{side} = 256$, which
splits the sphere into 786432 pixels.
This number of pixels corresponds to a 13.5' by 13.5' pixel on the sphere, and
fits well the scanning strategies we simulate.
Indeed, using $N_{side} = 512$ pixels would leave many holes in the maps and
therefore harm their visibility and further analysis.
To pixelize correctly the maps of a given experiment, and do the further
analysis, one needs of course to
use pixels two or three times smaller than the beam diameter.
Maps of many modern experiments would be well pixelized with pixels a bit
smaller than we use, but it would mean large maps
to store and a bit longer computation times (as the convergence speeds of the
methods would be lower), without giving any extra result to these
investigations.
In this article, we do not deal with the beam of an experiment and
we do not treat the issue of deconvolving the beam effect.

\subsection{Pixel averaging}
Reprojecting timelines on maps is not only a domain to domain
transform, but the simplest way to estimate the true map of the sky,
by averaging the samples of a same pixel on the sky. The noise is
therefore reduced by a factor square root of the number of samples in the
pixel (the weight).
To increase the signal-to-noise ratio in the pixels, one has to increase the counts
in a pixel, so increase the pixel size.
However the size pixel limits
of course the maximum l to be investigated ($l_{max}=2.N_{side}$).
If we consider pixel sizes bigger than the beam, or neglect the beam effect, then the $A^t$ matrix is
adding samples in a pixel, which means that this matrix is filled with 1
and 0 only.
This is much easier to handle than the true point-spread matrix, as in this
simple case the map-making (pixel averaging) of a timeline y is just $[A^tA]^{-1}A^t y$, where
$A^tA$ is a hit counter per map pixel.
A is in this case just the projection from a map to a timeline.
We will not investigate the beam deconvolution in this article, and therefore
consider only 1-and-0 point-spread matrices.
We present in Fig. 3 the weight maps for the three experiment strategies
that we have described.
It is clear that the 24 h balloon-borne polar flight has the best redundancy.
We will see if these differences in the experiment strategy produce important
differences in the final reconstructed maps.

\begin{figure}
 \begin{center}{
   \epsfxsize 10.0 true cm
    \leavevmode
  }\end{center}\caption[]{Weight maps: polar flight on top, equatorial flight
    in the center, satellite at bottom.
The units are in number of timeline samples in the pixel.}
\end{figure}

\subsection{Correlation properties}
The optimal map-making methods use the noise and sky covariance matrices N and
S.
These are impossible to invert or even to store for such large
timelines.
However, if the noise is stationary in the time domain, as it is the case for 1/f
noise and white noise (usual bolometer noises), the noise correlation matrix
in the time domain is circulant, i.e. multiplying a vector by this matrix is a
convolution, which is
filtering in the Fourier domain.
The noise correlation matrix has to be a statistical average on the
realizations of the $nn^t$ matrix, which means that a fit has to be
performed on the Fourier noise vector, to remove the gaussian fluctuations.
The Fourier noise vector can be estimated from the data,
since the 1/f tail is usually visible.
We fit this noise power spectrum with the $1/f^n$ law expressed below.
For the sky covariance matrix, the problem is different.
The sky covariance matrix is stationary in the map domain, as far as the Cosmic
Microwave Background is a gaussian random field.
Thus the S matrix (in the map domain) is circulant for CMB maps.
However, it is more complicated for the sky covariance matrix in the
timeline.
The sky covariance matrix in the time domain is stationary if two
conditions are verified:
the sky covariance matrix has to be stationary in the map domain,
and the sampling on the sky has to be constant, that means that the scanning
speed has to be constant if the sampling frequency is constant, which is
usually the case.
If the last
condition is not verified, then the stationarity of the CMB signal is
lost in time domain.
However, the tolerance threshold to this condition has to be investigated.
If we consider both the sky timeline and the noise timeline to be
stationary, then multiplying by $S^{-1}$ or $N^{-1}$ matrices is a convolution,
i.e. a vector multiplication in the Fourier domain.

\subsection{Direct and iterative methods}
We present here three optimal vector-only map-making methods: a COBE iterative
method, a Wiener direct method and a Wiener iterative method.

\subsubsection{COBE iterative method}
The COBE equation (Eq. 3) cannot be directly applied with vector-only algorithms,
because of the matrix inversions needed.
Thus the trick is to solve rather:

$[A^{t} N^{-1} A]$ \~x = $A^{t} N^{-1}$ y

This form prevents from the heavy inversion, but needs an iterative scheme.
The general iterative scheme for this equation is:

$\alpha$ \~x$_{n+1}$ = $\alpha$ \~x$_n$ + $A^{t} N^{-1}$ y - $[A^{t} N^{-1} A]$ \~x$_n$

where $\alpha$ is any linear operator on a vector, that is, any square matrix.
We have tested this algorithm on simulations and real data from the Archeops
experiment \cite{benoit01}, with $\alpha$ being a scalar.
By testing the method with different $\alpha$, we find that the identity is
the best iterator.

Another scheme can be developed, by making the noise map converge instead of
the sky map, as mentioned by Prunet (2001).
This can be better, as the signal can be more tricky than the instrumental
noise for the stability of the iterative scheme: hot galactic points for example
may induce stripes on the maps.
The noise-iterating scheme works with the following trick:
we change the variable \~x to \v{x} = $[A^tA]^{-1} A^t$ y - \~x.
It is straightforward to show that this is the noise map plus the
reconstruction error.
It leads to:

$\alpha$ \v{x}$_{n+1}$ = $\alpha$ \v{x}$_n$ + $A^{t} N^{-1}$ z - $[A^{t} N^{-1} A]$ \v{x}$_n$

where z = $A[A^tA]^{-1}A^t$ y - y.
If this algorithm converges, then the converging limit is exactly the optimal
solution of the map-making problem: the proof is that if it converges, then
\v{x}$_{n+1}$ = \v{x}$_n$ at the limit, and so the above equation leads to
the map-making equation for \v{x}$_n$, which is therefore the unique solution.
Of course, this true solution is
different from the true (simulated) sky, the residual noise covariance matrix
being $[A^tN^{-1}A]^{-1}$ in the case of the COBE method.
The case of $\alpha$ = I is actually the simplest iterator one can imagine,
but works well on the simulations and real data that we have processed.
We present in Section 5 the maps made with the iterative noise-converging COBE
method.

\subsubsection{Wiener direct method}
The Wiener 1 matrix (Eq. 5) is reduced to the Wiener filter \cite{wiener49} in the time domain, if few
conditions are verified.
Indeed, the Wiener 1 equation can be written as:

$W = [A^tA]^{-1} A^t ASA^{t} [ASA^{t} + N]^{-1}$

$ASA^{t}$ is the sky covariance matrix in the timeline, and N is the noise
covariance matrix in the timeline.
Thus if both the noise and the sky are stationary in the time domain, then the
Wiener 1 method is mapping the filtered timeline, with the optimal Wiener
filter being $\sigma^2\over{\sigma^2+\nu^2}$ in the Fourier domain, with 
$\sigma$ and $\nu$ the fitted Fourier transforms of the sky and the noise.
It is important to notice that two conditions are necessary so that the signal
is stationary in the timeline: the signal has to be stationary in the map, and
the observation has to be stationary towards the map, which means that both the
scanning speed and the sampling frequency have to be constant.
The noise Fourier fit is made as described for the COBE iterative method.
The sky power spectrum in the timeline is dominated by the rotation peaks, and
so to approximate the realization average by the power spectrum itself is not
a bad approximation.
This can be easily done even for real data by simulating the observation of a
pure sky (without noise).
We present in Fig. 4 the power spectra of the simulated noise timeline and the
simulated sky timeline.

\begin{figure}
 \begin{center}{
   \epsfxsize 9.0 true cm
    \leavevmode
  }\end{center}\caption[]{Power spectra of the polar flight noise timeline
    (black) and the sky timeline (blue).
The rotation peaks on the sky power spectrum are clearly visible.}
\end{figure}

The Wiener direct method is optimal when applied to stationary gaussian random signals,
but we test it also on non-gaussian and non-stationary skies, the first being
non-stationary even in the map domain because of the Galaxy, the other being
non-stationary on the timeline because of large variations of the scanning
speed, and we present the results in Section 5.

\subsubsection{Wiener iterative method}
The Wiener 2 matrix (Eq. 4), although identical to the Wiener 1, is computationally
nearer to the COBE matrix (Eq. 3).
Thus the iterative scheme is very close to the COBE one:

$\alpha$ \v{x}$_{n+1}$ = $\alpha$ \v{x}$_n$ + u - $[S^{-1} + A^{t} N^{-1} A]$
\v{x}$_n$

where u = $A^tN^{-1}[A[A^tA]^{-1}A^ty - y] + S^{-1}[A^tA]^{-1}A^ty$.
As we have shown for the COBE iterative method, the converging limit is the
exact solution of the Wiener map-making equation.
This iterative scheme needs to handle both the N matrix, noise covariance
matrix in the timeline, that we process as a filter in the Fourier domain
like we do for the COBE iterative method, and the S matrix, sky covariance
matrix in the map domain.
Handling this as a matrix is not possible for a small scale pixelization that
we need for CMB experiments of today, thus we have to process it as a filter
in Fourier space, like we do for filtering timelines.
The HEALPix RING scheme is stationary with respect to the sphere, because it
pixelizes it making a ring around the sphere from the north pole to the south
pole, with equal pixel surfaces.
So filtering a HEALPix vector (i.e. a map) in Fourier space is optimal, to the
condition that there must not be large holes in the map, that would harm the
stationarity of the sky in the HEALPix scheme.
Of course this means that for maps like the polar flight ones or the
equatorial flight ones, the filtering in the HEALPix scheme has to be made
separatly for each observed part of the sky, following the HEALPix ordering.
Another method is to define another pixelization, adapted to a given
observation.
But our satellite whole sky survey experiment is particularly well adapted to
test this method, as the holes are few in the $N_{side}$ = 256 pixelization,
so we can just filter the whole map.
We present the resulting maps in Section 5.

\section{Results and discussion}

\subsection{Testing the COBE iterative method and the Wiener direct method on
  the polar flight data}

We have made 1000 iterations with the COBE method, for the
polar simulated data containing the CMB fluctuations, the dipole, the Galaxy and the instrumental
noise.
We present the convergence plot of the rms of the difference map between the
reconstructed sky and the true sky (residual noise map) in Fig. 5 (top).
The level of white noise is shown as a blue straight line at 21.1 $\uk_{CMB}$
rms.
At the bottom of Fig. 5, we present the convergence plot made by considering
at each iteration the less well reconstructed pixel of the map, i.e. the
maximum of the residual noise map.

\begin{figure}
 \begin{center}{
   \epsfxsize 9.0 true cm
    \leavevmode
  }\end{center}\caption[]{Top: evolution of the difference map rms between the reconstructed sky
  and the true simulated sky.
The level of white noise is shown as a blue straight line at 21.1 $\uk_{CMB}$.
Bottom: evolution of the maximum of the residual noise map.
}
\end{figure}

The method reaches the residual noise at about 50 iterations, and this
residual is about 21.6 $\uk_{CMB}$ rms.
We can check that we have reached the convergence by observing the evolution
of the global residual noise rms, but also the evolution for some individual pixels, the map
aspect and the $C_l$ power spectrum.
We have to compare this result to the white noise amount in the map.
We can calculate the white noise in the map by two ways: one is to calculate
it theoretically as $n_{rms}^2 . \Sigma_i w_i^{-1}/N_{pix}$, where $n_{rms}$
is the white noise rms level in the timeline, $w_i$ the weight in pixel i, and
$N_{pix}$ the total number of observed pixels (the sum is of course on the
observed pixels).
The other is to simulate a white noise timeline and
make a map.
The rms level of white noise in the polar flight map is 21.06 $\uk_{CMB}$ rms,
which is very close to the residual noise amount.
This shows how good the reconstruction is, as it is clear that the correlated
noise is significantly removed from the map.
The reconstructed map exhibits no visible difference with the true map (whole
simulated sky
in Fig. 1).
We present in Fig. 6 the residual noise map.
The noise feature which can be seen in this map is the granularity of the white
noise, which confirms that the iterative method removes nearly completely the
correlated noise.
Moreover, the bottom plot in Fig. 5 shows that there are no pixels excluded
from the general convergence.
We present in Fig. 7 the $C_l$ power spectra of the map made by simple
pixel-averaging, of the residual noise
map after convergence, and of a
map made with only white noise at the same level.
All spectra are corrected from the fraction of the sky covered.
The units show very well how better is the iterative method, compared to the
pixel averaging.
The noise spectrum is nearly the one of a white noise above about l=50, but
exhibits some weak residual correlation at larger scales.
Since we can be confident in the fact that the convergence has been reached,
then the map obtained has to be the solution of the map-making problem.
We see that a small amount of correlated noise still lurks at large scales:
this may be explained by the fact that these scanning strategies do not
constrain perfectly the largest scales (few crossing scans), which are the most noisy (by
definition of the 1/f noise).
Of course, a method such as MASTER (Hivon \etal 2001, subm. to \apj) may be able to estimate this
large-scale noise excess.
However, this correlated residual noise amount is extremely low, even at the largest
scales, and could have to be taken into account only for extremely precise
measurements of the low l (precision better than 0.05 $\uk ^2$ on the $C_l$ at
large scales !).

\begin{figure}
 \begin{center}{
   \epsfxsize 10.0 true cm
    \leavevmode
  }\end{center}\caption[]{Residual noise map obtained with the COBE iterative method applied on the polar flight data.
The much largest part of this residual is clearly the averaged white noise.
}
\end{figure}

\begin{figure}
 \begin{center}{
   \epsfxsize 9.0 true cm
    \leavevmode
  }\end{center}\caption[]{Top: $C_l$ power spectrum of the polar flight
residual noise map
made by pixel averaging.
Bottom: $C_l$ power spectra of the residual noise map after convergence
(black), and of a white noise map (blue).
All spectra are corrected from the fraction of the sky covered.
The units show very well how better is the iterative method, compared to the
pixel averaging.
The noise spectrum is nearly the one of a white noise above about l=50, but
exhibits some weak residual correlation at larger scales.
}
\end{figure}

We have also tested the method on timelines made with just CMB fluctuations and
instrumental noise.
To get a timeline without the Galaxy from real data, it is possible, for
instance, to decorrelate with a shorter wavelength bolometer.
Indeed, most CMB experiments have high frequency channels (for
instance Planck and Archeops bands at 353 and 545 GHz) in which the
CMB fluctuations are completely negligible. These bands are used as
tracers of the dust Galactic emission.
The way to decorrelate the CMB bands from the dust emission is usually
to do it after the map-making process (component separation), but it
is also possible to decorrelate the Galactic emission in the
timelines, though it may cause noise propagation in the CMB-band
timeline.
The right way to do it seems to us (though it is not the purpose of
the paper) to get first a Galactic map as clean as possible (from some
high frequency bolometers, or even from standard Galactic maps as IRAS
ones...),
and then re-sample it with the scanning strategy of the timeline one
wants to decorrelate.
But we have to note also that many CMB experiments do not observe the
Galaxy at all, and thus have directly nearly CMB-only timelines.

The residual noise obtained on the reconstructed CMB-only map is the same as
the one obtained from complete data.
This shows that the COBE iterative method is as efficient with different sky
signals, as we could expect from the theory.
This algorithm is very efficient, but quite slow:
the full CPU time on a 500 MHz PC is about one hour per 25 iterations, for the 8 million
sample timeline.
However, we shall see hereafter that the number of iterations needed to reach
the convergence can be much reduced, if we use a first estimator for the sky
map.
Also, it is possible to speed up this kind of iterative methods with
multi-grid algorithms \cite{dore01}.

\begin{table}
\caption[]{\label{table: rms}
This table presents the noise amounts in the maps made with data containing
the CMB fluctuations and the instrumental noise.
The first line shows the noise rms in the coadded maps, the second shows the
rms of the white (uncorrelated) noise, the third shows the residual noise rms
after convergence of the iterative COBE method, the fourth after the iterative
Wiener method.
The last line shows the residual noise rms after the Wiener direct filtering.
The units are in $\uk_{CMB}$, and
n.c.s.s. stands for non constant scanning speed.

}
\begin{flushleft}
\begin{tabular}{lllll}

\hline
 & Polar & Polar 24 h &
 Equat. & Satel. \\
 & 24 h & n.c.s.s. & 12 h & \\
\hline
\hline

{\small Total noise rms} & 349.7 & 352.5 & 498.6 & 378.4\\
{\small in the map} & & & & \\
\hline

{\small White noise rms} & 21.06 & 21.37 & 40.84 & 38.14\\
{\small in the map} & & & & \\
\hline

{\small Residual rms} & 21.61 & - & 42.13 & 38.93\\
{\small COBE iter.} & & & & \\
\hline

{\small Residual rms} & - & - & - & 38.93\\
{\small Wiener iter.} & & & & \\
\hline

{\small Residual rms} & 22.82 & 33.82 & 32.34 & 41.40\\
{\small Wiener direct} & & & & \\
\hline

\end{tabular}
\end{flushleft}
\end{table}

We have applied the Wiener 1 direct method to the same timeline.
This method is particularly well adapted to CMB-only signals, as the
stationarity is perfectly verified, and more fundamentaly because the Wiener
method assumes a gaussian prior for both the sky and the noise.
The direct Wiener method works very well on the CMB-only data: the difference map between the
Wiener direct reconstructed map and the true sky is 22.8 $\uk_{CMB}$ in
rms, which is close to the white noise level.
One could wonder what a "wild" whitening filter used instead of the optimal
Wiener filter could produce: we present in Fig. 8 the map made with a
whitening filter, that is the inverse of the fitted noise power spectrum.
The result is that the signal is clearly destroyed by this non-optimal filter
(compare with the CMB map in Fig.1).

\begin{figure}
 \begin{center}{
   \epsfxsize 10.0 true cm
    \leavevmode
  }\end{center}\caption[]{CMB map constructed with a whitening filter, from the polar flight data.
The signal is clearly destroyed by this non-optimal filter.
}
\end{figure}

We present in Fig. 9 the residual noise map obtained with the Wiener direct
method.

\begin{figure}
 \begin{center}{
   \epsfxsize 10.0 true cm
    \leavevmode
  }\end{center}\caption[]{Residual noise map obtained with the Wiener direct
    method applied on the polar flight data.
There is some weak residual striping, to the contrary to what is obtained
after iterations.
}
\end{figure}

Some weak residual striping is still visible in this map.
It shows that the Wiener direct filtering processes the data in a very
different way than does the COBE iterative method.
The iterative method removes gradually the correlated noise, to end with a
nearly perfectly destriped map, whereas the Wiener direct filter processes the
data in order to properly construct the sky map.
In this method, it is the signal which is restored, and not the correlated
noise which is removed.
This may explain why some correlated noise can be in the final map while the
total residual noise amount is very weak.
It is important to keep in mind that not only the quantity of residual noise
is important, but above all its "whitness": if the residual noise is quite
correlated, then the pixel noise matrix in the reconstructed map is not
diagonal, which is a problem for the $C_l$ estimation.
The $C_l$ spectrum of the Wiener-direct-reconstructed map shows that some
power is suppressed at all scales, whereas it is not the case for the COBE
iterative method, for which the $C_l$ power spectrum exhibits no significant
difference with the true CMB power spectrum.

An idea to accelerate the convergence of the iterative methods is to use a non-zero
first estimate of the map, the best being the Wiener direct map.
It is important to understand that whatever is the first estimate, what
defines the exact solution is only the map-making equation,
thus the Wiener 1 first estimate used for COBE iterations accelerates the
convergence of the COBE iterative method, but does not influence the final
reconstructed map.

In order to test how the lack of stationarity in the signal can harm the Wiener
direct filtering, we test the method on a timeline made with CMB and Galactic
signals, plus the instrumental noise.
We also test it on a CMB timeline suffering from an extremely non constant
scanning speed, the beam doing 30 degrees amplitude swingings 17 times per minute.
The Galactic Wiener direct map is quite badly reconstructed, since stripes are
still visible and the residual noise is 257 $\uk_{CMB}$ in rms.
This is not surprising, as the intense Galactic signal harms very clearly the
stationarity of the sky signal.
The Wiener direct filtering is thus not adapted to timelines containing
Galactic signals.
It may therefore be applied to timelines whose foregrounds have been removed
first, or to observations made away from the Galactic plane.
To the contrary, the Wiener direct map is well reconstructed for the data made
with the very non constant scanning speed: the residual noise is 33.8
$\uk_{CMB}$ in rms, which is not bad compared to the 21.4 $\uk_{CMB}$ rms of
white noise in the map, but of course clearly less good than the 22.8 $\uk_{CMB}$ rms of
residual noise obtained with the same processing on constant scanning speed data.
However, the good result obtained with these dreadfully sampled data shows
the robustness of the method.

\subsection{Comparing the different observations toward the map-making methods}

We have applied the COBE iterative method with the Wiener 1 direct map as
first estimate, to the equatorial flight and satellite data, with CMB
fluctuations and instrumental noise.
Once again, this can also be applied to signals including the Galaxy, but
since the Wiener direct estimate would not be so good, the number of
iterations needed to get a perfectly destriped map would be higher.
The maps made with the COBE iterative method for these two other experiments
are very well reconstructed too: for the equatorial flight simulations, the
residual noise map is 42.1 $\uk_{CMB}$ in rms, with 40.8 $\uk_{CMB}$ rms of
white noise.
For the satellite data, the residual noise map is 38.9 $\uk_{CMB}$ rms, with
38.1 $\uk_{CMB}$ rms of white noise.
The residuals of the iterative COBE method are close to the white noise
amount, and the residual over white noise ratios are nearly constant.
Of course the residual is larger for the low-redundancy experiments, but the
efficiency of the iterative method is not less.
In particular, this shows that numerous crossing scans are not crucial for the map-making process.

It is interesting to notice that in the case of the equatorial flight, the
Wiener direct filtering gives a lower residual than the COBE iterative method,
and even lower than the white noise amount.
However, the residual noise map (Fig. 10) exhibits some residual striping.

\begin{figure}
 \begin{center}{
   \epsfxsize 10.0 true cm
    \leavevmode
  }\end{center}\caption[]{Residual noise map obtained with the Wiener direct method applied on the equatorial flight data.
There is some residual striping, even if the total residual noise amount
is extremely weak.
}
\end{figure}

\subsection{Testing the Wiener iterative method on the satellite simulated data}
As explained before, a whole sky survey, even with a few holes in the map, is
a perfect candidate to test the Wiener iterative method.
The method works well even with no first estimate of the sky, but is much
faster with a Wiener 1 first estimate.
We present in Fig. 11 the residual noise map of the Wiener iterative method.
This map is very close to the one of the COBE iterative method (no visible
difference, and very close rms: 38.933 $\uk_{CMB}$ for
the COBE iterative method, 38.931 $\uk_{CMB}$ for the Wiener iterative method).
Once again, we observe that the correlated noise is nearly totally removed by the iterations.
The Wiener method is known to minimize the reconstruction error: we check it,
but the difference with the COBE method is very small.
Moreover, the $C_l$ power spectra of the true CMB sky and both reconstructed
maps (COBE iterative and Wiener iterative) do not exhibit significant differences.

\begin{figure}
 \begin{center}{
   \epsfxsize 10.0 true cm
    \leavevmode
  }\end{center}\caption[]{Residual noise map obtained with the Wiener iterative method applied on the satellite data.
The granularity of the white noise is clearly visible.
}
\end{figure}

\subsection{Generalization of the iterative map-making methods}

The noise-converging iterative scheme can be generalized to balance the
influence of the {\it a priori} signal correlation knowledge in the map-making
process.
This is done with the following map-making matrix:

$W = [\epsilon S^{-1} + A^{t} N^{-1} A]^{-1} A^{t} N^{-1}$ \hspace*{1em} (Eq. 6, Saskatoon)

It is the Saskatoon way to make maps (Tegmark \etal 1997 and
Tegmark 1997), where the $\epsilon$ factor allows to choose the signal-to-noise ratio in the final map.
Actually, the Wiener map-making gives generally less noisy maps than the COBE
one, by reducing the power in the pixels unequally.
This induces a given signal-to-noise ratio in the final map, which is optimal
with respect to the minimization of the reconstruction error, in the case of
the Wiener method.
As the noise is more important at small scales, the Wiener map-making smoothes
the map in an adequate way.
However, it is interesting to assume a different signal-to-noise ratio, in
order, for instance, to try to retrieve small scale features in the map.
For $\epsilon = 0.5$, we obtain a residual noise map of 38.932 $\uk_{CMB}$ in
rms, that we could expect.
The method is unstable for $\epsilon > 1$.

We have seen that the difference exists much more between the iterative methods
and the direct ones, than between COBE iterative and Wiener iterative.
Indeed, the iterative methods are very precise, both toward the map aspect and
the $C_l$ spectrum, and do not seem to allow significant differences between
Wiener and COBE, with this quite good signal-to-noise ratio.
Though, we can be confident in the reality of the Wiener iterative
method (compared to the COBE one) because the epsilon factor is very
sensitive (when superior to 1, the method diverges).

\section{Conclusion}
We have presented efficient map-making methods for large Cosmic Microwave
Background timelines.
The direct way to estimate the sky signal from a noisy timeline is the Wiener
direct filtering, which is robust towards scanning strategy non-stationary
effects, but needs to process a stationary signal, which demand to remove the
Galactic signal before map-making.
The iterative methods (COBE, Wiener, and generalized) are very efficient to
remove the correlated stationary noises, and allow to get a very good precision on the
reconstructed map, whose residual noise is nearly only averaged white noise.
Though the results of COBE, Wiener and "Saskatoon" iterative methods are very close, we
check that the Wiener estimation more reduces the global reconstruction error.
The difference should be larger if the pixels are more unequally noisy: it
seems that this kind of experiment strategies does not allow the
Wiener estimation to be very different from the {\it no prior} one.
The different scanning strategies that we have simulated allow as efficient
map-making, even if the crossing scans are few.
However, it is important to keep in mind that we have processed stationary
noises, which means a timeline free from, for instance, periodic systematics
which could be projected on the maps like features on the sky.

This kind of vector-only methods seems to us unavoidable to make optimal maps
from CMB experiments of today, or still to come.
The reduction of the information in CMB data is a heavy work, from gigabytes of
rawdata to essentially 12 cosmological numbers with their error bars.
Since the computer facilities are limited and unsufficient for brute force
approaches (and it will be still the case for Planck data reduction), it is
an interesting challenge to process each step of this reduction work without
losing information.
Using stationarity properties of a signal in a given domain (sphere, map,
timeline...) to transform a matrix inversion problem into a vector-only
solution, could be probably also developed for other CMB reduction steps, such
as the component separation.

\section{Acknowledgements}
We would like to thank K.M. G\'orski and his collaborators for their so
useful HEALPix package.
We also thank very much F. Bouchet, F.-X. D\'esert, O. Dor\'e, P. Filliatre,
J.-C. Hamilton, and the whole Archeops collaboration.

\end{document}